\documentclass[aps,reprint,superscriptaddress,floatfix,showpacs,pra]{revtex4-2}
\usepackage{graphicx}
\usepackage{amsmath}
\usepackage{amssymb}
\usepackage{amsfonts}
\usepackage{appendix}

\usepackage[unicode=true,breaklinks=true,colorlinks=true]{hyperref}
\hypersetup{citecolor=blue,urlcolor=blue}

\usepackage{newtxtext}
\usepackage{newtxmath}
\usepackage[all]{hypcap}
\usepackage{natbib}
\usepackage{multirow}
\usepackage{booktabs}
\usepackage{bbold}
\usepackage{bm}
\usepackage{bbm}

\newcommand{\ket}[1]{|{#1}\rangle}
\newcommand{\bra}[1]{\langle{#1}|}

\begin{document}
	
	\title{Rapid transform optimization strategy for decoherence-protected quantum register in diamond}
	\author{Jiazhao Tian}
	\email{tianjiazhao@tyut.edu.cn}
	\affiliation{School of physics, Taiyuan university of technology, Taiyuan 430000 P. R. China}
	\author{Haibin Liu}
	\affiliation{School of physics, Hubei University, Wuhan 430062 P. R. China}
	\author{Roberto Sailer}
	\affiliation{Institute for Quantum Optics and Center for Integrated Quantum Science and Technology, Ulm University, D-89081 Ulm, Germany}
	\author{Liantuan Xiao}
	\email{xiaoliantuan@tyut.edu.cn}
	\affiliation{School of physics, Taiyuan university of technology, Taiyuan 430000 P. R. China}
	
	\author{Fedor Jelezko}
	\affiliation{Institute for Quantum Optics and Center for Integrated Quantum Science and Technology, Ulm University, D-89081 Ulm, Germany}
	\author{Ressa S. Said}
	\email{ressa.said@uni-ulm.de}
	\affiliation{Institute for Quantum Optics and Center for Integrated Quantum Science and Technology, Ulm University, D-89081 Ulm, Germany}

	\begin{abstract}
		Decoherence-protected spins associated with nitrogen-vacancy color centers in diamond possess remarkable long coherence time, which make them one of the most promising and robust quantum registers. The current demand is to explore practical rapid control strategies for preparing and manipulating such register. Our work provides all-microwave control strategies optimized using multiple optimization methods to significantly reduce the processing time by $80\%$ with a set of smooth near-zero-endpoints control fields that are shown to be experimentally realizable. Furthermore, we optimize and analyze the robustness of these strategies under frequency and amplitude imperfections of the control fields, during which process we use only $16$ samples to give a fair estimation of the robustness map with $2500$ pixels. Overall, we provide a ready-to-implement recipe to facilitate high-performance information processing via decoherence-protected quantum register for future quantum technology applications.
	\end{abstract}
	
	\maketitle

	\section{Introduction}
	Nitrogen-vacancy color centers (NV center) in diamond are important candidate for future quantum computer registers as well as enhanced quantum sensors\cite{aslamNanoscaleNuclearMagnetic2017, zaiserEnhancingQuantumSensing2016} because of their long coherence time. Utilizing decoherence-free subspace (DFS) of the nuclear spin systems\cite{zhouCavityQEDImplementation2015, yunParallelpathImplementationNonadiabatic2022, ramakoteswararaoLevelAnticrossingsNitrogenvacancy2020a} can even prolong their coherence time and strengthen their advantages in future quantum technologies. A wide range of systems coded in DFS have been studied in areas of quantum computing and quantum sensing\cite{kwiatExperimentalVerificationDecoherencefree2000, wuCreatingDecoherencefreeSubspaces2002}, include atoms\cite{kockumDecoherenceFreeInteractionGiant2018a}, trapped ions\cite{kielpinskiDecoherenceFreeQuantumMemory2001, wuHolonomicQuantumComputation2005}, solid state quantum dots\cite{taylorSolidStateCircuitSpin2005} and so on\cite{reillyAdiabaticControlDecoherencefree2022, suHeraldedQuantumEntangling2023, huOptimizingQuantumGates2021}.  Recently, a DFS strategy based on one NV$^-$ center and two nearby nuclear spins has been proposed, showing high resistance against static noises of $\sigma_z$ direction\cite{gonzalezDecoherenceprotectedQuantumRegister2022}. In this strategy, the preparation and manipulation of the DFS are based on stimulated Raman adiabatic passage (STIRAP) driven by microwave control field, avoiding the slow process of directly driving nuclear spins through radio frequency field. Nevertheless, the total transition time of the STIRAP strategy is still limited by the adiabatic condition. A method of Superadiabatic STIRAP (SA-STIRAP)\cite{chenShortcutAdiabaticPassage2010} can efficiently speed up the velocity by one order of magnitude, provided that an auxiliary radio frequency field is applied. However, the required strength of this radio field is in magnitude of megahertz, making it a challenge to faithfully achieve the field in experiment due to the nonlinear relationship between the Rabi frequency and the drive amplitude\cite{sangtawesinHyperfineenhancedGyromagneticRatio2016}.
	
	In this work we present the all-microwave control optimization strategies to construct and manipulate nuclear spins in DFS of NV-based nuclear system, decreasing the evolution time by one order of magnitude while maintaining the same fidelity. The control fields have smooth shapes and near-zero values at the beginning and ending points, making them feasible to be realized in experiments. As a recipe for optimization strategies in real experiments, we compare three common optimization methods, namely the gradient-based Gradient Ascent Pulse Engineering (GRAPE) method\cite{khanejaOptimalControlCoupled2005}, the multivariate function optimization methods Chopped Random Basis (CRAB) method\cite{canevaChoppedRandombasisQuantum2011} and the Phase Modulation (PM) method\cite{tianQuantumOptimalControl2020} in terms of the optimization result and speed under different evolution time. We show the simulated and experimental field shapes given by each method, confirming the experimental feasibility of these strategies. As frequency and amplitude bias and noise of the control field are inevitable in practice, we further make a fast estimation as well as optimization of the field robustness using the Bayesian esticonfirmingmation phase-modulated (B-PM) method\cite{tian_bayesian-based_2023}. With this method only $16$ samples are adequate to give the $50 \times 50$ pixel fidelity distribution map. In general, we supply versatile, fast and realistic toolbox of optimization methods to facilitate further implementations of NV center as quantum register and sensor. 
	
	The structure of the paper is arranged as follows. In section \ref{section1} we introduce the construction and manipulation strategy based on the decoherence-protected space of the NV-nuclear system. Section \ref{section_optimization} represents optimization results of the GRAPE, CRAB and PM method in terms of fidelity, optimization speed and field shape. Section \ref{section_experimental feasible} investigates and analyses the experimental feasibility of the optimization fields in terms of real control shape given by control apparatuses (arbitrary wave generator and amplifier), as well as the robustness under frequency and amplitude bias, where we optimize this robustness using B-PM method. In section \ref{section_discussion} we provides summary and discussions of our results.
	
	\section{Decoherence-protected space of  nuclear spins}\label{section1}
	
	The system under consideration is a tri-partite system comprising one NV$^-$ electron spin ($S = 1$) and two proximal $^{13}$C nuclear spins ($I = 1/2$). The effect of near $^{14}$N nuclear spins can be eliminated by polarization technology in experiment\cite{hansonPolarizationReadoutCoupled2006}. The total Hamiltonian of the system reads\cite{gonzalezDecoherenceprotectedQuantumRegister2022}
	
	\begin{equation}
		H=D \hat{S}_{z}^{2}+\gamma_{e} \mathbf{B} \cdot \mathbf{S}+\mathbf{S} \cdot \sum_{i=1}^{2} \mathbb{A}^{(i)} \cdot \mathbf{I}^{(i)}+\gamma_{c} \mathbf{B} \cdot \sum_{i=1}^{2} \mathbf{I}^{(i)}+H_{n n},
	\end{equation}
	where $D \hat{S}_{z}^{2}$ is the zero field term of the electron spin, $\gamma_{e} \mathbf{B} \cdot \mathbf{S}$ is the magnetic interaction of the electron spin, $\mathbf{S} \cdot \sum_{i=1}^{2} \mathbb{A}^{(i)} \cdot \mathbf{I}^{(i)}$ is the hyperfine coupling of the electron spin and the nuclear spins, $\gamma_{c} \mathbf{B} \cdot \sum_{i=1}^{2} \mathbf{I}^{(i)}$ is the magnetic interaction of the nuclear spin, and $H_{n n}$ is the dipole coupling between nuclear spins. The explicit forms of the hyperfine coupling tensor $\mathbb{A}^{(i)}$ for $i$th nuclear spin and $H_{n n}$, as well as the detail procedure to simplify the Hamiltonian can be found in Appendix \ref{AppSec: H}. Hereafter, we use the simplified form,	   
	\begin{equation}\label{Eq:H}
		H = 
		\begin{bmatrix}
			H^{m_{s}=0} & 0 \\
			0 & H^{m_{s}=1} 
		\end{bmatrix},
	\end{equation}
	where $H^{m_{s}=0}$ and $H^{m_{s}=1}$ is the $4 \times 4$ subspace Hamiltonian of two nuclear spins when the electron spin number is $m_{s} = 0$ and $m_{s} = 1$ respectively, with the explicit form
	\begin{equation}\label{H_ms=0}
		\begin{aligned}
			H^{m_{s}=0}=\gamma_{c} B_{z}\left(\hat{I}_{z}^{(1)}+\hat{I}_{z}^{(2)}\right)+\gamma_{c} B_{x}\left(\hat{I}_{x}^{(1)}+\hat{I}_{x}^{(2)}\right)\\
			+\frac{d_{12}}{2}\left(\left(\hat{I}_{+}^{(1)} \hat{I}_{-}^{(2)}+\hat{I}_{-}^{(1)} \hat{I}_{+}^{(2)}\right)-4 \hat{I}_{z}^{(1)} \hat{I}_{z}^{(2)}\right)
		\end{aligned}
	\end{equation}
	and
	\begin{equation}\label{H_ms=1}
		H^{m_{s}=1}=\left(D+\gamma_{e} B_{z}\right)\mathbb{1}+A_{z z}^{(1)} \hat{I}_{z}^{(1)}+A_{z z}^{(2)} \hat{I}_{z}^{(2)}
	\end{equation}
	respectively, where $\mathbb{1}$ represents the $4\times4$ identity matrix. The value of parameters in Eq.(\ref{H_ms=0}) and Eq.(\ref{H_ms=1}) are shown in Table \ref{tab:param_value_in_total_Hamil}.
	\begin{table*}[]
		\centering
		\caption{Parameter values in the total Hamiltonian\cite{gonzalezDecoherenceprotectedQuantumRegister2022}\normalsize}
		\label{tab:param_value_in_total_Hamil}
		\begin{tabular}{lll|lll}
			\hline
			parameter & symbol & value & parameter name & symbol & value \\
			\hline
			zero-field splitting&    $D$    & $2\pi\times2.87$ GHz      & dipolar coupling &  $d_{12}$      &   $4$ kHz        \\
			gyromagnetic ratio of electron spin &   $\gamma_{e}$     &  $2\pi\times 2.8$ MHz/G     &     isotropic hyperfine coupling      &  $A_{zz}^{(1)}$      &  $12.45$ MHz      \\
			gyromagnetic ratio of nuclear spin&  $\gamma_c$      &  $2\pi\times 1.7$ kHz/G    & isotropic hyperfine coupling    &  $A_{zz}^{(2)}$    & $2.28$ MHz  \\
			\hline
		\end{tabular}
	\end{table*}
	\begin{figure*}[] \label{Fig_Hamiltonian}
		\centering 
		\includegraphics[width = 17cm]{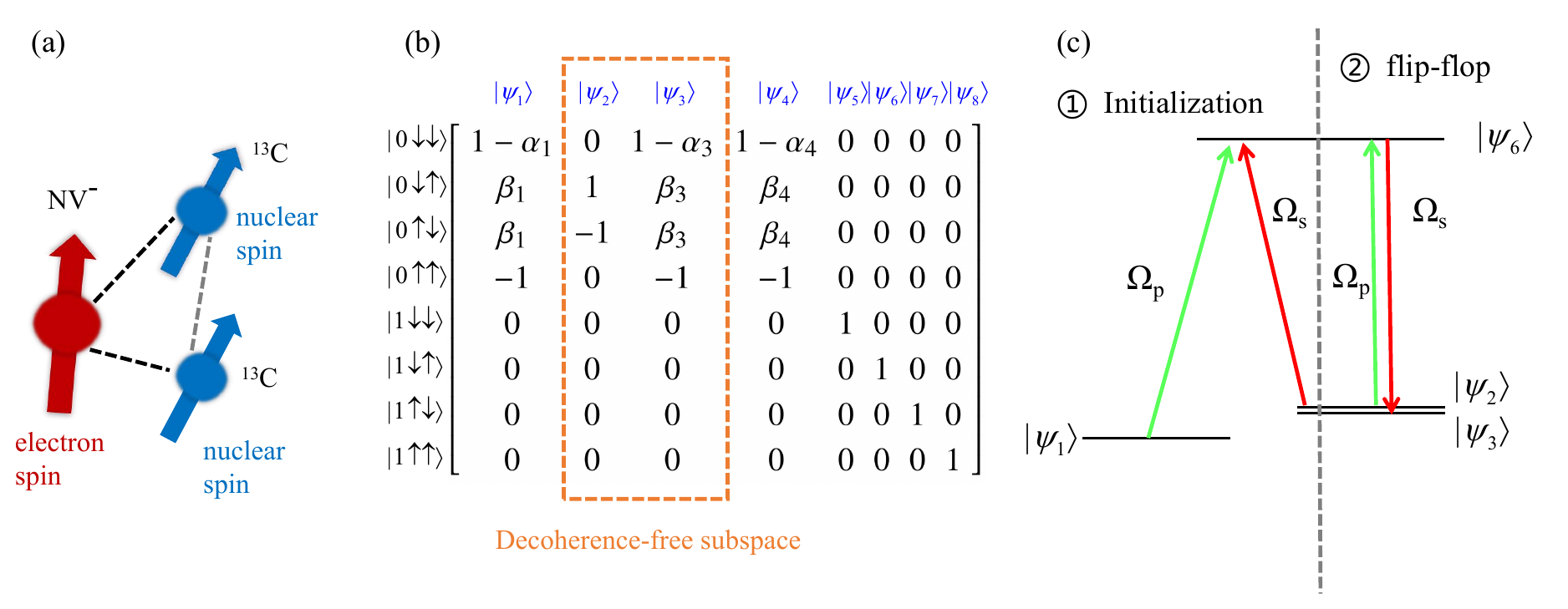}
		\caption{(a) The tri-particle system comprising one NV$^-$ electron spin ($S = 1$) and two proximal $^{13}$C nuclear spins ($I = 1/2$). (b) Explicit formation of eigenstates (without normalization) of Hamiltonian of the tri-particle system showed in Equation (\ref{Eq:H}). For $j = 1,3,4$, $\alpha_{j} = \frac{(2 d_{12}-2 E_j)(-d_{12}-2 E _j-2 B_z \gamma_{c})}{2 B_x^2 \gamma_ c^2}$, $\beta_{j} = \frac{-d_{12}-2 E_j-2 B_z \gamma_{c}}{2 B_x \gamma_c}$, $E_j$ being the solution of the equation $-2 d_{12}^3-4 B_x^2 d_{12} \gamma_c^2+8 B_z^2 d_{12} \gamma_c^2-\left(3 d_{12}^2+4 B_x^2 \gamma_c^2+4 B_z^2 \gamma_c^2\right) E_j+E_j^3 = 0$. The explicit values of $E_1$, $E_3$ and $E_4$ under different magnetic fields are shown in Appendix \ref{AppSec: H}. (c) Schematic diagram of the transition process $\ket{\psi_{1}}\rightarrow\ket{\psi_{2}}$ in the initialization step and $\ket{\psi_{2}}\rightarrow\ket{\psi_{3}}$ in the spin flip step.}
	\end{figure*}
	Denoting the eigenstates and corresponding eigenvalues of $H$ as $|\psi_{i}\rangle$ and $E_i$ respectively ($i = 1,2,\cdots 8$), $H$
	can be represented by $H=\sum_{i} E_{i} \hat{\sigma}_{ii}$, where $\hat{\sigma}_{i j}=\left|\psi_{i}\right\rangle\left\langle\psi_{j}\right|$. The explicit forms of $|\psi_{i}\rangle$ are showed in Figure \ref{Fig_Hamiltonian} (b).
	Specifically, $|\psi_{2}\rangle$ has zero eigenvalue, $|\psi_{3}\rangle$ has near-zero eigenvalue (see caption of Figure \ref{Fig_Hamiltonian} (b) and Appendix \ref{AppSec: H} for details), which makes both of these states robust against fluctuations of magnetic field and possess long dephasing time. The states $|\psi_{2}\rangle$ and $|\psi_{3}\rangle$, therefore, construct the decoherence-protected subspace.
	
	To process logical qubit operations in the decoherence-protected subspace based on $|\psi_{2}\rangle$ and $|\psi_{3}\rangle$, one should firstly initialize the system into $|\psi_{2}\rangle$ or $|\psi_{3}\rangle$, then carry out flip-flop between these two logic states. The initialization process follows the transitions $\ket{0\uparrow\uparrow}\rightarrow\ket{\psi_{1}}\rightarrow\ket{\psi_{2}}$, where the first step $\ket{0\uparrow\uparrow}\rightarrow\ket{\psi_{1}}$ can be completed by tuning the amplitude of magnetic field in a time scale of $20$ $\mu$s\cite{gonzalezDecoherenceprotectedQuantumRegister2022} (see Appendix \ref{AppSec: Initialization}). The second step of the initialization process, $\ket{\psi_{1}}\rightarrow\ket{\psi_{2}}$, as well as the flip-flop process $\ket{\psi_{2}}\leftrightarrow\ket{\psi_{3}}$, need to be driven by an external control field. Direct driving with radio frequency field are typically slow due to the low gyromagnetic ratios of the nuclear spin \cite{PhysRevLett.124.220501}. Although, in principle increasing the intensity of the control field can speed up the driving process, in realistic experiments the dynamics of the nuclear spin oscillations become nonsinusoidal under strong control field \cite{sangtawesinHyperfineenhancedGyromagneticRatio2016}.
	Indirect control with microwave fields using $m_s = 1$ states $\ket{\psi_{5,6,7,8}}$ as an intermediate circumvents this issue and achieve rapid transition\cite{PhysRevLett.124.220501}. 
	
	Taking the transition from $\ket{\psi_{1}}$ to $\ket{\psi_{2}}$ for example, we write the driving Hamiltonian as
	\begin{equation}
		H_{\text{d}} = \left(\sqrt{2} \Omega_{p} \cos \left(\omega_{p} t\right)+\sqrt{2} \Omega_{s} \cos \left(\omega_{s} t\right)\right)  \hat{S}_{x},
	\end{equation}
	where $\Omega_{p(s)}$ and $\omega_{p(s)}$ are the amplitudes and the frequencies of two MW fields with $\omega_{p} = E_6-E_1$, $\omega_{s} = E_6-E_2$, and $\hat{S}_{x} = \left(|1\rangle\langle 0|+|0\rangle\langle 1|\right)/\sqrt{2}$. For brevity, we use the interaction Hamiltonian with respect to
	\begin{equation}
		H_0 = E_1\sigma_{11} + E_2\sigma_{22} + E_6\sum_{i=5}^{8}\sigma_{ii}.
	\end{equation}
	After neglecting rapid oscillation terms we obtain the interaction Hamiltonian with rotating-wave approximation,
	\begin{equation}\label{Eq:RWA_Hamiltonian}
		\begin{aligned}
			H^I_{\text{RWA}}=&\frac{\Omega_{p}(t)}{2}\left(\chi_{51} \hat{\sigma}_{51}+\chi_{61} \hat{\sigma}_{61}+\chi_{71} \hat{\sigma}_{71}+\chi_{81} \hat{\sigma}_{81}+\text { h.c. }\right)\\
			&+\frac{\Omega_{s}(t)}{2}\left(\chi_{62} \hat{\sigma}_{62}+\chi_{72} \hat{\sigma}_{72}+\text { h.c. }\right)+(H-H_0),
		\end{aligned}
	\end{equation}
	with $\chi_{i j}=\left\langle\psi_{i}|\hat{V}| \psi_{j}\right\rangle$, and $\hat{V}=\sqrt{2}\hat{S}_{x}$. The detailed derivation can be found in Appendix \ref{AppSec: Hd}. Considering transverse relaxations, evolution of the tri-particle system can be described by the Lindblad master equation
	\begin{equation}\label{Eq:masterEq}
		\frac{\mathrm{d} \rho(t)}{\mathrm{d} t}=-i\left[H^{I}, \rho\right]+\mathcal{L}_S+\mathcal{L}_I,
	\end{equation}
	where $\mathcal{L}_S = \left(1 / T_2^*\right)\left(2 \hat{S}_z \rho \hat{S}_z-\hat{S}_z^2 \rho-\rho \hat{S}_z^2\right)$ is the dissipation term of electron spin and $\mathcal{L}_I = \sum_{i=1}^2\left(1 / T_{2 n_i}^*\right)\left(2 \hat{I}_z^{(i)} \rho \hat{I}_z^{(i)}-\hat{I}_z^{(i)^2}\rho-\rho \hat{I}_z^{(i)^2}\right)$ is the dissipation term of nuclear spins. Here, the coherence times are taken as $T_2^*=7 \mu \mathrm{s}, T_{2 n_1}^*=500 \mu \mathrm{s}$ and $T_{2 n_2}^*=700 \mu \mathrm{s}$\cite{gonzalezDecoherenceprotectedQuantumRegister2022}. The transition effectiveness under certain evolution time $T$ can be measured by the fidelity between final density matrix $\rho(T)$ and the target density matrix $\ket{\psi_{2}}\bra{\psi_{2}}$, represented by \cite{khanejaOptimalControlCoupled2005}
	\begin{equation}\label{Eq:F}
		F = \text{Tr}(\bra{\psi_{2}}\rho(T)\ket{\psi_{2}}). 
	\end{equation}
	
\begin{figure*}[] 
		\centering 
		\includegraphics[width = 16cm]{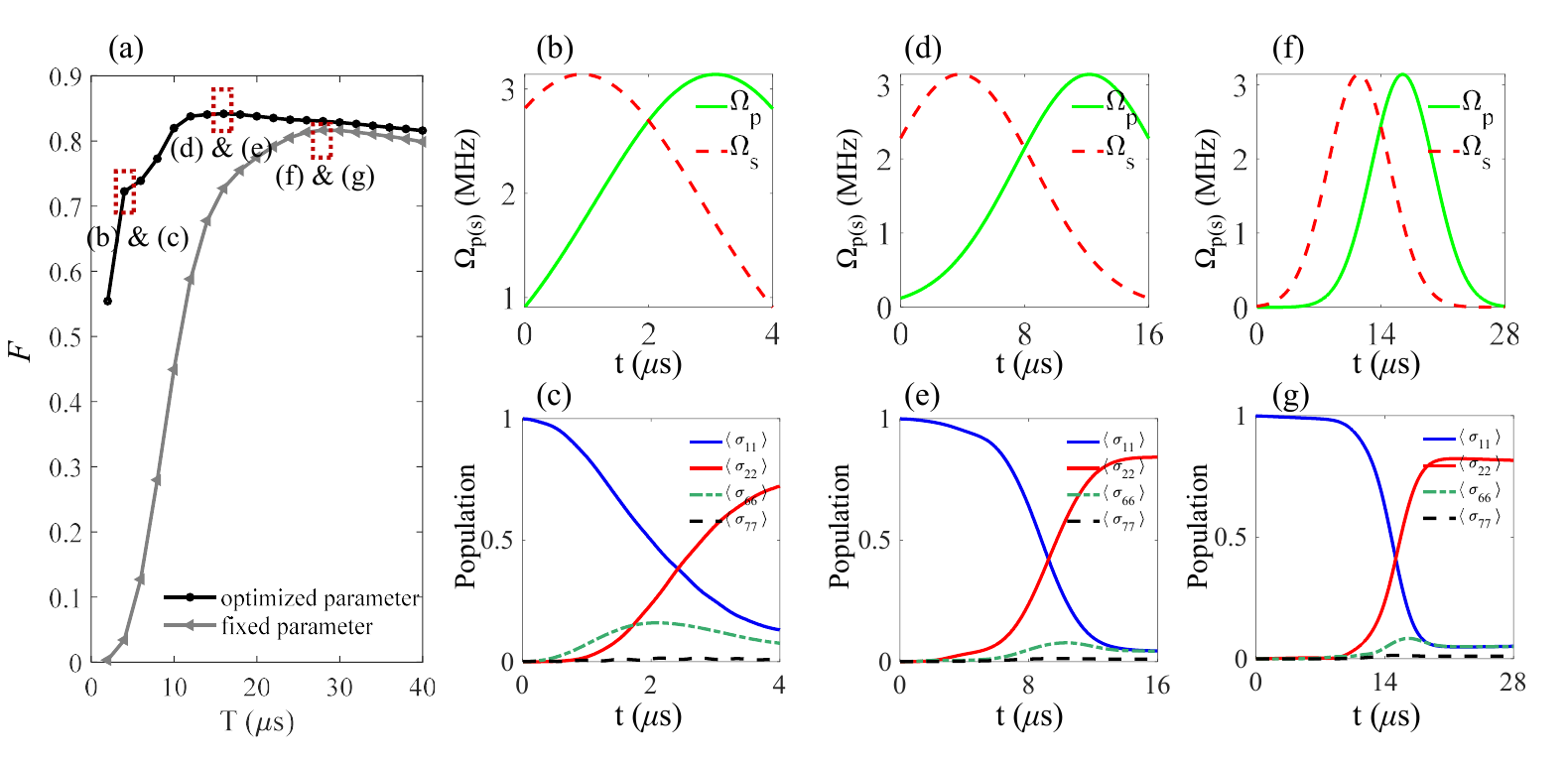}
		\caption{(a) Transition efficiency of the process $\ket{\psi_{1}} \rightarrow \ket{\psi_{2}} $ given by STIRAP methods with Gaussian shape control field following Equation (\ref{Eq: Gauss field}). Black line with dots: results with optimized parameters $\sigma$ and $t_{\mathrm{d}}$. Gray line with triangles: results with fixed parameters $\sigma = T/8$ and $t_{\mathrm{d}} = \sqrt{2}\sigma$.
			(b,c) Control field and population transition at $T = 4$ $\mu$s, with optimized parameter $\sigma = 1.95$ $\mu$s and $t_{\mathrm{d}} = 2.15$ $\mu$s. (d,e) Control field and population transition at $T = 16$ $\mu$s, with optimized parameter $\sigma = 4.77$ $\mu$s and $t_{\mathrm{d}} = 8.34$ $\mu$s. (f,g) Control field and population transition at $T = 28$ $\mu$s, with fixed parameter $\sigma = T/8 = 3.5$ $\mu$s and $t_{\mathrm{d}} =  \sqrt{2}\sigma = 4.95$ $\mu$s. \label{Fig_STIRAP_Opt} }
	\end{figure*}

	One conventional strategy to complete the state transition with high fidelity is the stimulated Raman adiabatic passage (STIRAP), where the amplitudes of  control fields take the Gaussian shape
	\begin{equation}\label{Eq: Gauss field}
		\begin{aligned}
			& \Omega_p(t)=\Omega_0 \exp \left(-\left(t-t_{\mathrm{d}} / 2\right)^2 / 2 \sigma^2\right), \\
			& \Omega_s(t)=\Omega_0 \exp \left(-\left(t+t_{\mathrm{d}} / 2\right)^2 / 2 \sigma^2\right).
		\end{aligned}
	\end{equation}
	As showed in Figure (\ref{Fig_STIRAP_Opt}) (a), this adiabatic transition (with $\sigma = T/8$, $t_{\mathrm{d}} = \sqrt{2}\sigma$) needs the evolution time being $T \geq 24$ $\mu$s to obtain fidelities of $F \geq 0.8$.  When $T$ decreases to $4 \mu$s, the fidelity drops to $F = 0.034$. We optimize the values of parameter $t_{\mathrm{d}}$ and $\sigma$ by a direct search method to obtain the highest fidelity under different evolution times, which increase the fidelity to $F = 0.842$ at $T = 16$ $\mu$s. However, fidelity at $T = 4$ $\mu$s is still as low as $F = 0.723$. In addition, as showed in Figure (\ref{Fig_STIRAP_Opt}) (b) and (d), the optimized control fields have beginning or ending values around $2\sim 3$ MHz. In practical experiments, such fields with non-zero endpoints could be distorted more severely due to the bandwidth limitation of the amplifier, thus decrease the final fidelity. Different strategies are essential to improve the results with higher fidelity and more practical field shape.

	\begin{figure*}[t] 
		\centering 
		\includegraphics[width = 16cm]{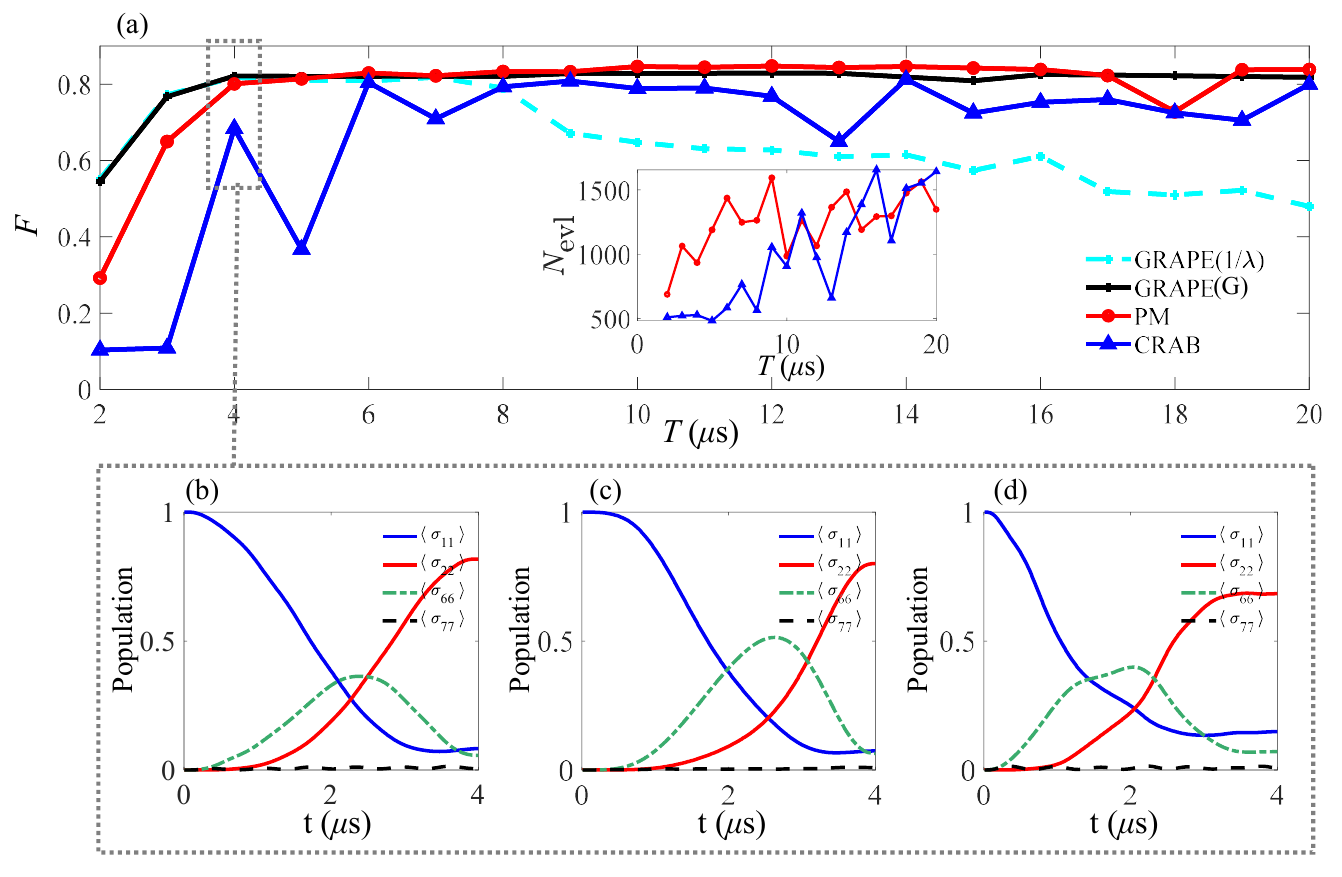}
		\caption{(a) Optimization fidelity with different optimization methods under different evolution time. Inserted figure: Average number of function evaluation of PM and CRAB method. (b-c) Population transition at $T = 4$ $\mu$s, given by GRAPE(G), PM and CRAB method respectively. \label{Fig_ThreeMethod}}
	\end{figure*}

	\begin{figure*}[h] 
		\centering 
		\includegraphics[width = 15cm]{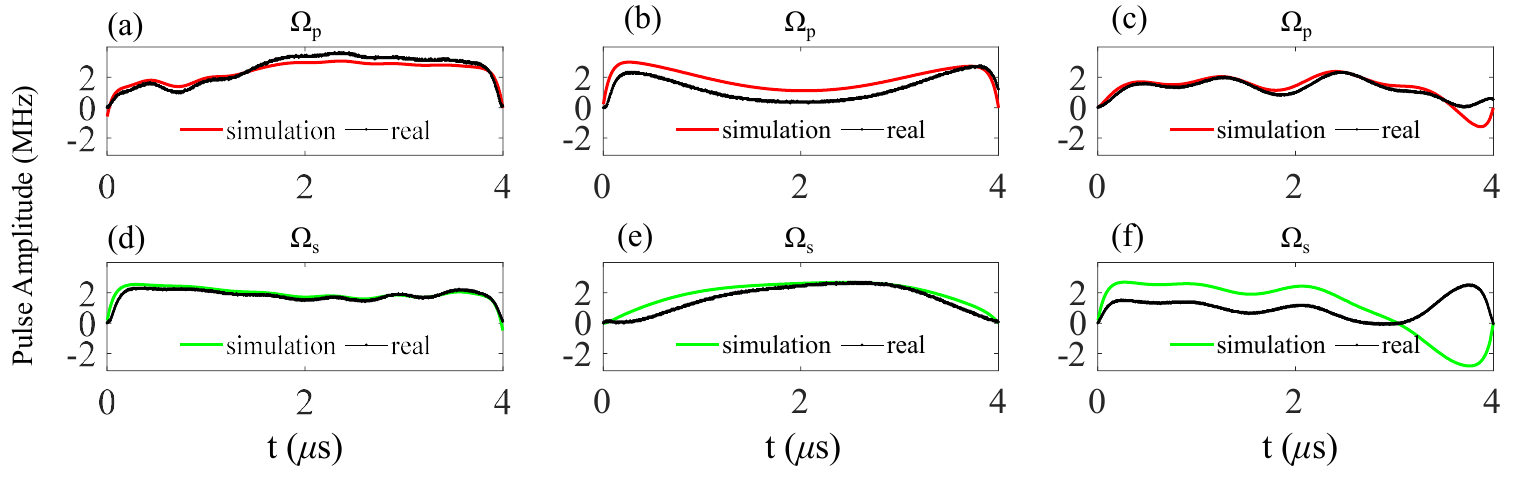}
		\caption{Comparison of the simulation pulse and the real pulse generated by the AWG (Tektronix AWG-70002A, connected with the software QUDI) and measured by the oscilloscope  (LECROY-WAVEACE 234) after a RF amplifier (Model 60S1G4AM3 AR Germany with frequency bandwidth 0.7-4.2 GHz and gain power of 60 W ) for the same gain level. The simulation pulses are given by different optimization method GRAPE (figure (a) and (d)), PM (figure (b) and (e)) and CRAB (figure (c) and (f)). Amplitudes of the measured values of the real pulses are scaled to make a direct comparison visible (see the main text for details).\label{Fig_true}}
	\end{figure*}
	
	\begin{figure*}[h] 
		\centering 
		\includegraphics[width = 12cm]{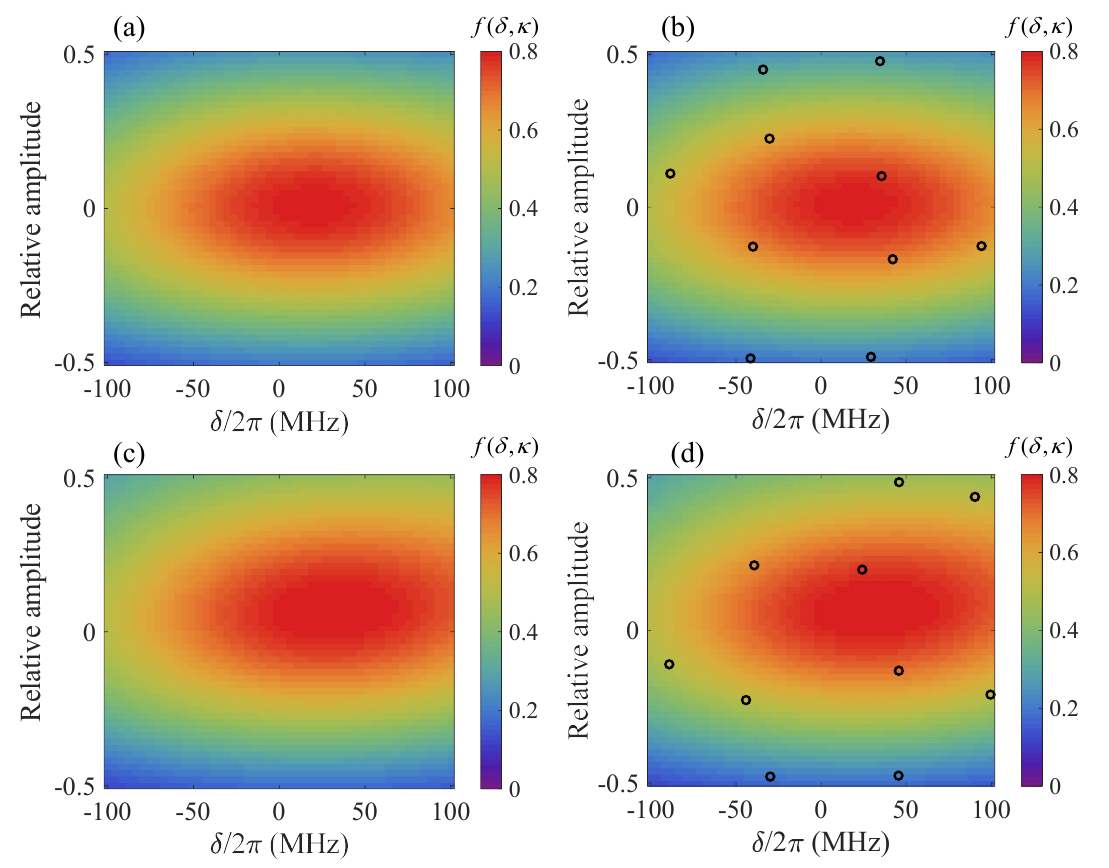}
		\caption{(a) Simulated fidelity of the transition process from $\ket{\psi_{1}}$ to $\ket{\psi_{2}}$ under different values of detuning and amplitude bias of the control field. The total evolution time is $T = 4$ $\mu$s, and the control field is given by the PM method, showed in Figure \ref{Fig_true} (b,e). (b) Estimation values of the fidelity using the Bayesian based estimation method. The black circles represent the location of samples. $16$ randomly chosen sample points are used, only those with locations within the range ($\delta/2\pi \in [-100,100]$ MHz, $\kappa \in [-0.5, 0.5]$) are showed. (c) Simulated fidelity of the control field optimized by the B-PM method. (d) Estimation values of the fidelity using the Bayesian based estimation method with the optimized field given by the B-PM method. The black circles represent the location of samples.  $16$ randomly chosen sample points are used, only those with locations within the range ($\delta/2\pi \in [-100,100]$ MHz, $\kappa \in [-0.5, 0.5]$) are showed.\label{Fig_robust} }
	\end{figure*}

	\section {Optimization}\label{section_optimization}
	In what follows we show how various optimization methods improve the low transition efficiency of STIRAP when the evolution time is reduced to several microseconds. We consider three widely used methods, the gradient ascent pulse engineering (GRAPE) method\cite{khanejaOptimalControlCoupled2005}, the chopped random basis (CRAB) method\cite{canevaChoppedRandombasisQuantum2011}, and the phase-modulated (PM) method\cite{tianQuantumOptimalControl2020}. GRAPE is an exploitative method, in the sense that it update each pulse section along the gradient ascending direction, and the value of its objective function will converge to a local optimum. In contrast, CRAB and PM are explorative based on global search method, where the expansion coefficients of the control fields are taken as the optimization parameters. Both methods use truncated expansion, while the CRAB method features randomization of the frequencies and the PM method features phase modulation formation to improve the optimization efficiency. 
	
	Considering the transition process from $\ket{\psi_{1}}$ to $\ket{\psi_{2}}$ and taking the fidelity $F$ in Equation (\ref{Eq:F}) as the objective function, all three methods give the optimal shapes of control fields $\Omega_p (t)$ and $\Omega_s (t)$ that maximize $F$ under the constraint $\Omega_{p(s)} (t) \leq \Omega_{\text{max}}$, where we set the maximum field amplitude as $\Omega_{\text{max}} = \pi$ MHz. We use a boundary function \cite{scheuer2014precise}  $\lambda(t)=h^p /\left(h^p-(t-h)^p\right)$ with $h = T/2$ and $p = 30$ to obtain near-zero value of starting and ending points, which makes the control fields more practical in experiments. The stopping criteria is set as the termination tolerance on function value being less than $10^{-4}$.
	
	For the GRAPE method, $\Omega_p$ and $\Omega_s$ are constructed by $N$ pulses respectively with equal width $\Delta T = T/N$, and the optimization parameters are the amplitudes of these $2N$ pulses represented by $u_j^{p}$ and $u_j^{s}$ respectively, with $j = 1, 2, \cdots, N$. Considering the local property of GRAPE method, we use two types of initial values to explore the optimums in a larger range. One is denoted as GRAPE ($1/\lambda$), with initial field $\Omega_{p(s)}(t) = \Omega_0/\lambda(t)$, where $\Omega_0$ is uniformly distributed random number in the range $[0.5\pi, 0.9\pi]$ MHz. Another is denoted as GRAPE (G) with the Gaussian shape initial field in Equation (\ref{Eq: Gauss field}) with $\sigma$ randomly taken from $[0.3, 3] \mu$s, $t_{\mathrm{d}} = \sqrt{2}\sigma$ and $\Omega_0 = 0.9 \pi$ MHz. In each interaction of the GRAPE process, these pulses are updated successively according to the format\cite{khanejaOptimalControlCoupled2005}
	\begin{equation}
		u_j^{p(s)} = u_j^{p(s)} + \epsilon \frac{\partial F}{\partial u_j^{p(s)}},
	\end{equation}
	where $\epsilon$ is the interaction step length, the value of which should be set properly to guarantee the convergence of $F$. Meanwhile, the explicit forms of control fields in CRAB method are
	\begin{equation}
		\Omega_p = \frac{1}{\lambda(t)}\sum_{n=1}^{N_c} \left( A_n \sin \left(\omega_n t\right)+B_n \cos \left(\omega_n t\right)\right),
	\end{equation}
	\begin{equation}
		\Omega_s = \frac{1}{\lambda(t)}\sum_{n=1}^{N_c} \left( C_n \sin \left(\omega_n t\right)+D_n \cos 	\left(\omega_n t\right)\right),
	\end{equation}
	with $\omega_n = 2\pi n (1+r_k)/T$, $r_k \in [-0.5,0.5]$ represent random
	numbers with flat distribution\cite{canevaChoppedRandombasisQuantum2011}.  
	For the PM method, the explicit forms of control fields are
	\begin{equation}
		\Omega_p = \frac{1}{\lambda(t)} \sum_{n=1}^{N_c} a_n\cos \left[\frac{b_n}{v_n} \cos \left(v_n t\right)\right],
	\end{equation}
	\begin{equation}
		\Omega_s = \frac{1}{\lambda(t)} \sum_{n=1}^{N_c} a_n\cos \left[\frac{b_n}{v_n} \sin \left(v_n 	t\right)\right].
	\end{equation}
	The optimization parameters of the CRAB and PM method are the $N_c\times 4$ matrix $[\boldsymbol{A},\boldsymbol{B},\boldsymbol{C},\boldsymbol{D}]$ and the $N_c\times 3$ matrix $[\boldsymbol{a},\boldsymbol{b},\boldsymbol{v}]$, respectively, where $\boldsymbol{A} = [A_1, A_2, \cdots, A_n]^{\prime}$, other vectors have similar forms. Here, we take $N_c = 3$, which corresponds to the fewer numbers of harmonics hence more friendly to control apparatuses.

	Figure \ref{Fig_ThreeMethod} (a) shows the optimization fidelity given by GRAPE, CRAB and PM methods under different evolution times. At $T = 4$ $\mu$s, GRAPE method with both kinds of initial values gives $F \approx  0.82$ ($0.82$ for GRAPE(G) and $0.817$ for GRAPE($1/\lambda$) explicitly) and PM method gives $F = 0.8$. CRAB method only reaches $F = 0.683$. The highest fidelity $0.848$ appears at $T = 11\quad \mu$s. Comparing results of different methods we see the GRAPE(G) method shows a stable performance at different time conditions, while the GRAPE($1/\lambda$) method perform bad for longer evolution time. The PM method is more possible to give the highest fidelity when $T \geq 7$ $\mu$s but can also gives bad results at a few time points, this instability indicates a lack of total trial numbers. The CRAB method fall behind in both aspects of highest fidelity and stability. Overall, the fidelities exceeds $0.8$ when $T \geq 4$ $\mu$s, indicating $4$ $\mu$s is the shortest time required to efficiently complete the transformation. The comparison of optimization speed of two direct search method, PM and CRAB, are showed in the inserted figure in Figure \ref{Fig_ThreeMethod} (a), evaluated by the average calling number of objective functions. In a time scale of  $T< 10$ $\mu$s, the CRAB method shows a speed advantage, and when $T> 10$ $\mu$s, CRAB and PM methods show similar behavior. Explicit population transition at $T = 4$ $\mu$s are shown in Figure \ref{Fig_ThreeMethod} (b-d). 
	
	\section {Experimental feasibility}\label{section_experimental feasible}
	The main concern on the experimental feasibility is whether the optimized field can be accurately realized in the experiments. To test and demonstrate such experimental feasibility, we generate the control fields using an arbitrary wave generator (Tektronix AWG-70002A, connected with the software QUDI) and measure the output electronical signal by an oscilloscope (LECROY-WAVEACE 234). The bandwidth of our AWG is $14$ GHz and the amplitude resolution is $10$ bits. We denote the optimized simulation pulse sequences by $u_{\text{sim}}$, which comprising $N=T/\Delta T$ flat pulses $u_{\text{sim}}(i)$ ($i = 1, 2, \cdots, N$) with pulse length of $\Delta T$. Similarly the measured amplitudes of the pulse sequences are denoted by $u_{\text{real}}$, which is comprising $u_{\text{real}}(i)$ ($i = 1, 2, \cdots, N$). To compare the shapes of the simulation and real pulses, we translate the values of measured signal to make sure they begin from zero, and scale them by a factor of $d_{\text{sim}}/d_{\text{real}}$, where $d_{\text{sim}} = \max |u_{\text{sim}}|-|u_{\text{sim}}(1)|$ is the difference between the maximal amplitude and amplitude of the beginning 
	pulse of the simulation control field, and $d_{\text{real}} = \max |u_{\text{real}}|-|u_{\text{real}}(1)|$ is the difference between the maximal amplitude and amplitude of the beginning pulse of the output pulse sequence. The comparison results are showed in Figure \ref{Fig_true}, where the real pulse shapes are broadly consistent with the simulation pulse shapes, which implies the optimization methods are indeed feasible in practice. Details of the experiment and the true values for conversion between voltage signal and Rabi frequency are given in the Appendix \ref{AppSec: Experiment}.

	In practical experiments, besides the limitation of available bandwidth of the arbitrary wave generator (AWG) and amplifier, another inevitable disturbance of the transition efficiency is the noise originating from ambient nuclear spins and external bias fields, which can be represented as the fluctuation of the amplitude as well as the detuning of the control field.  Using the optimal PM control field showed in Figures \ref{Fig_true} (b) and \ref{Fig_true} (e), we calculate the fidelity for different values of detuning and amplitude bias of the control field, the results are showed in Figure \ref{Fig_robust} (a). Such distribution map containing $50\times50$ pixels, requiring $2500$ calculation times of the evolution function. Using the Bayesian-estimation method\cite{tian_bayesian-based_2023}, we can significantly reduce this number from $2500$ to $16$, and get a fair estimation of the $50\times50$ pixel distribution map, showed in  Figure \ref{Fig_robust} (b). Based on this method, to improve the robustness of the control field, we further made an optimization using the Bayesian estimation phase-modulated (B-PM) method \cite{tian_bayesian-based_2023} with the objective function defined as the average fidelity
	\begin{equation}\label{Eq:F_ave}
		\mathcal{F}_{\mathrm{obj}}=\mathcal{N} \sum_{k=1}^M \sum_{j=1}^N p\left(\delta_k\right) p\left(\kappa_j\right) f\left(\delta_k, \kappa_j\right),
	\end{equation}
	where $f\left(\delta_k, \kappa_j\right)$ is the fidelity with control field detuning $\delta_k$ and amplitude bias $\kappa_k$ (see Appendix \ref{AppSec: H_delta}  for details), $p(\delta)$ and $p(\kappa)$ are the normal distribution of $\delta$ and $\kappa$:
	\begin{equation}
		\begin{aligned}
			&p(\delta)=\frac{1}{\sqrt{2 \pi} \sigma_\delta} e^{-\frac{\delta^2}{2 \sigma_\delta^2}}\\
			&p(\kappa)=\frac{1}{\sqrt{2 \pi} \sigma_\kappa} e^{-\frac{\kappa^2}{2 \sigma_\kappa^2}},
		\end{aligned}
	\end{equation}
	and $\mathcal{N}=\left[\sum_{k=1}^M \sum_{j=1}^N p\left(\delta_k\right) p\left(\kappa_j\right)\right]^{-1}$ is the normalization constant. The optimization result are presented in Figure \ref{Fig_robust} (c), and the estimation of Figure \ref{Fig_robust} (c) using $16$ samples are given in Figure \ref{Fig_robust} (d), which are visibly identical.
	
	\section{Discussions}\label{section_discussion}
	We have presented a comprehensive comparison of three wildly used methods, namely the GRAPE method, the CRAB method, and the PM method, based on the optimization fidelity, speed and experimental feasibility. Synthetically, we find the GRAPE method performing well with high fidelity and rapid optimization speed in shorter evolution time range, while the PM method shows stable performance for all evolution time within the scope of consideration and is easy-to-use as it can be accomplished by direct searching method. Besides, we achieve a fast and accurate estimation as well as optimization of the field robustness using the B-PM method. These results provide reference for existing STIRAP shortcut methods in the presence of dissipation\cite{stefanatosOptimalShortcutsStimulated2022,evangelakosOptimalSTIRAPShortcuts2023} as well as robust control methods of NV center\cite{oshnikRobustMagnetometrySingle2022,poulsenOptimalControlNitrogenvacancy2022a} to further explore and complete each other.
	
	Further optimizations can be carried out based on these methods, including optimization of the magnetic field amplitudes during the preparation and manipulation process of the system, since the bias of magnetic field is the common interference factor in typical experiments. One can also apply the methods to closed-loop control that directly uses experimental outputs as the objective function value for exploring more practical control field during the experimental process\cite{fengGradientbasedClosedloopQuantum2018,chenClosedLoopRobustControl2013,kochQuantumOptimalControl2022}. More advance investigations and analyses 
	on how the control apparatuses implement the numerical optimized pulses, such as frequency response analysis, will significantly improve the experimental performance of the optimized pulses \cite{singhCompensatingNonlinearDistortions2023}. The system under consideration could be expanded to scalable multiqubit registers in NV centers \cite{PhysRevX.9.031045,PhysRevX.9.031045,mailePerformanceQuantumRegisters2023}. Furthermore, our work can straightforwardly be adapted to an 
	alternative system involving an intrinsic Nitrogen nuclear spin\cite{zhangEfficientImplementationQuantum2020a}. Overall, we provide a versatile optimization strategy for improving performance of quantum register based on DFS nuclear spin systems in diamond for future quantum computing and sensing technologies.

	\begin{acknowledgments}
		JZT acknowledge supports from Science Foundation for Youths of Shanxi Province (202203021222113) and National Natural Science Foundation of China (62305241), and thanks Jianming Cai, Kangze Li, Yaoxing Bian, Jiamin Li and Shuangping Han for their discussions. XLT and JZT acknowledge supports from National Natural Science Foundation of China (U23A20380). FJ and RSS acknowledge supports from the DFG, BMBF (CoGeQ, SPINNING, QRX), QC-4-BW, Center for Integrated Quantum Science and Technology (IQST), and the ERC Synergy Grant HyperQ. RSS thanks Philipp Vetter (Ulm), Matthias Müller (FZ Jülich), and Phila Rembold (TU Wien) for their discussions. All of the authors are grateful to Jingfu Zhang (Ulm) for his advices and Timo Joas (Ulm) for providing codes in the pulse measurement process.
	\end{acknowledgments}
	
	\appendix
	
	\section{Static Hamiltonian}\label{AppSec: H}
	
	The system under consideration is a tri-partite system comprising one~NV$^-$ electron spin~($S = 1$), and two proximal $^{13}$C nuclear spins~($I = 1/2$)~\cite{gonzalezDecoherenceprotectedQuantumRegister2022}. The total time-independent Hamiltonian is\cite{gonzalezDecoherenceprotectedQuantumRegister2022}
	\begin{equation}
		H=D \hat{S}_{z}^{2}+\gamma_{e} \mathbf{B} \cdot \mathbf{S}+\mathbf{S} \cdot \sum_{i=1}^{2} \mathbb{A}^{(i)} \cdot \mathbf{I}^{(i)}+\gamma_{c} \mathbf{B} \cdot \sum_{i=1}^{2} \mathbf{I}^{(i)}+H_{n n},
	\end{equation}
	where~$D \hat{S}_{z}^{2}$ is the zero field term of the electron spin, $\gamma_{e} \mathbf{B} \cdot \mathbf{S}$ is the magnetic interaction of the electron spin, $\mathbf{S} \cdot \sum_{i=1}^{2} \mathbb{A}^{(i)} \cdot \mathbf{I}^{(i)}$ is the hyperfine coupling of the electron spin and the nuclear spins, while the hyperfine coupling tensor~$\mathbb{A}^{(i)}$ for $i$th nuclear spin has the form of 
	\begin{equation}
		A_{k l}^{(i)}=A_{c}^{(i)} \delta_{k l}+A_{d}^{(i)}\left(\delta_{k l}-3 \hat{r}_{k}^{(i)} \hat{r}_{l}^{(i)}\right),
	\end{equation}
	with the Fermi constant term~$A_c$, the dipolar interaction $A_d$, and~$k, l=x, y, z$. The term $\gamma_{c} \mathbf{B} \cdot \sum_{i=1}^{2} \mathbf{I}^{(i)}$ represents the magnetic interaction of the nuclear spin. The last term follows
	\begin{equation}\label{total_Hamiltonian}
		H_{n n}=\sum_{i<j} \frac{\mu_{0} \gamma_{c}^{2}}{4 \pi r_{i j}^{3}}\left(\mathbf{I}_{i} \cdot \mathbf{I}_{j}-\frac{3\left(\mathbf{I}_{i} \cdot \mathbf{r}_{i j}\right)\left(\mathbf{r}_{i j} \cdot \mathbf{I}_{j}\right)}{r_{i j}^{2}}\right).
	\end{equation}

	To achieve Hamiltonian of Equation (\ref{Eq:H}) in the main text, several approximations need to be made. Firstly, negligible quantity can be omitted, including: (1) The non-axial components, $S_x$ and $S_y$, are neglected due to $D \gg \gamma_{e} B_{x}$ and $D \gg A_{i j}$; (2) The isotropic hyperfine couplings are much stronger than the dipolar coupling ($d_{12} = 4$ kHz), as well as the magnetic interaction terms~$\gamma_{c}Bx$ and~$\gamma_{c}Bz$, therefore the later three terms vanish in the~$m_s = 1$ Hamiltonian. Besides, we consider a simplified spatial arrangement, where the direction vector $\mathbf{r}$ between the two nuclear spins are assumed to be parallel with the quantization axis~$\hat{z}$ and the magnetic field comprises only~$z$ and~$x$ directions, such that~$\mathbf{B}=B_{z} \hat{z}+B_{x} \hat{x}$. Further, for a more concise form of the Hamiltonian as well as the system description, the anisotropic hyperfine coupling $A_{ani}$ is neglected. However, using the full Hamiltonian does not change the main conclusions of the strategy based on the decoherence-free subspace\cite{gonzalezDecoherenceprotectedQuantumRegister2022}. 
	
\begin{figure*}[t] 
		\centering 
		\includegraphics[width = 15cm]{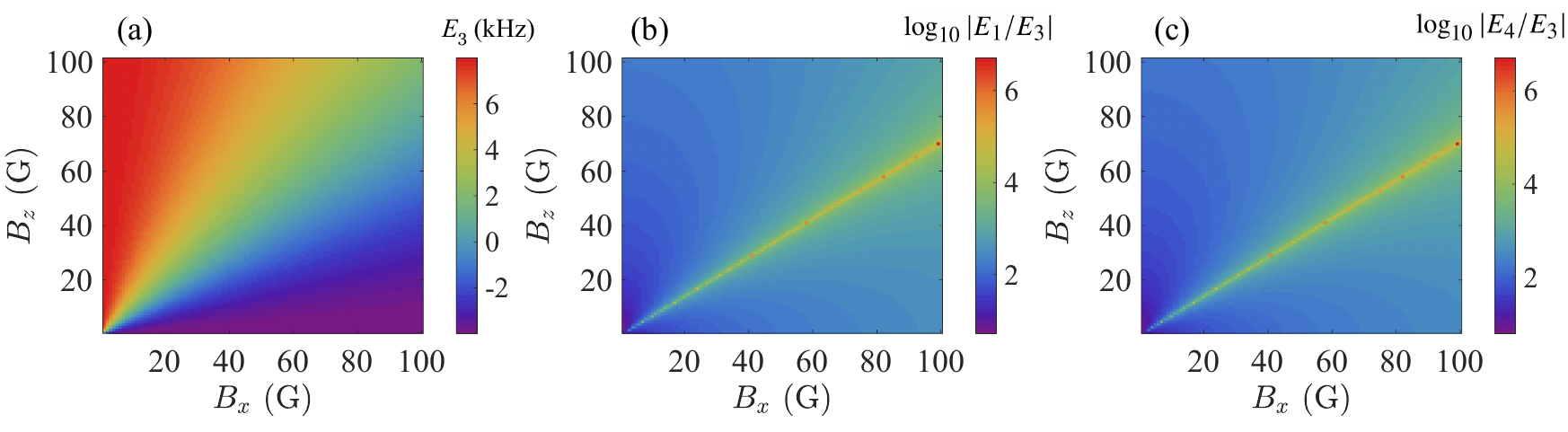}
		\caption{(a) Explicit value of $E_3$ under different $B_x$ and $B_z$ values. (b) Comparison between $E_1$ and $E_3$, evaluated by $\log_{10}|E_1/E_3|$. (c) Comparison between $E_4$ and $E_3$, evaluated by $\log_{10}|E_4/E_3|$. }
		\label{fig: E_value}
	\end{figure*}

	The explicit formation of eigenstates of the final Hamiltonian of Equation (\ref{Eq:H}) are showed in Figure \ref{Fig_Hamiltonian} (b) in main text. Figure \ref{fig: E_value} gives the explicate value of $E_3$, as well as comparisons between $E_{1(4)}$ and $E_3$ when the value of $B_x$ and $B_z$ varies from $1$ G to $100$ G, indicates in this range $E_3$ is indeed close to zero compared to $E_1$ and $E_4$.
	
	\section{Initialization process}\label{AppSec: Initialization}
	The strategy adopted to access the decoherence-protected subspace is by polarizing the spin system to the state~$\left|0_{L}\right\rangle$, that consists of three steps~\cite{gonzalezDecoherenceprotectedQuantumRegister2022}. The first step is to polarize the system to the state~$\left|0\uparrow\uparrow\right\rangle$, via the application of magnetic field having the components~$B_z\approx 5$G and $B_x\approx 70$G. With this magnetic field, the eigenstate~$\left|\psi_{1}\right\rangle$, is closely projected to the state~$\left|0\uparrow\uparrow\right\rangle$. Hence, the system is now approximately polarized to the state~$\left|\psi_{1}\right\rangle$. The second step is successively tuning~$B_x$ and~$B_z$, during which the state~$\left|\psi_{1}\right\rangle$ evolves adiabatically to a state that has balanced contribution of all bare states, reached when the magnetic field arrives at~$B_x = 100$G and $B_z = 5$G. The last step is the transformation from the state~$\left|\psi_{1}\right\rangle$ to $\left|\psi_{2}\right\rangle$, by applying microwave~(MW) fields that drive the transitions~$\left|\psi_{1}\right\rangle \rightarrow\left|\psi_{6}\right\rangle \rightarrow\left|\psi_{2}\right\rangle$. 
	
	\section{Driving Hamiltonian}\label{AppSec: Hd}
	In the following, we elaborate the derivation of the driving Hamiltonian~\cite{gonzalezDecoherenceprotectedQuantumRegister2022}, to the total system Hamiltonian, 
	\begin{equation}
		H=\sum_{i} E_{i} \hat{\sigma}_{ii},
	\end{equation}
	where~$\hat{\sigma}_{i j}=\left|\psi_{i}\right\rangle\left\langle\psi_{j}\right|$, with~$\left|\psi_{i}\right\rangle$ are the eigenstates~$(i=1, \ldots, 8)$, and~$E_i$ represents the corresponding eigenvalues. Our static system is driven by two MW fields, indexed by~$p$ and~$s$, giving the driving Hamiltonian as 
	\begin{equation}
		H_{\text{d}} = \left(\sqrt{2} \Omega_{p} \cos \left(\omega_{p} t\right)+\sqrt{2} \Omega_{s} \cos \left(\omega_{s} t\right)\right)  \hat{S}_{x},
	\end{equation}
	with~$\Omega_{p(s)}$ and $\omega_{p(s)}$ being the amplitudes and the frequencies of the two MW fields, respectively, and $\hat{S}_x=\left(|1\rangle\langle 0|+|0\rangle\langle 1|\right)/\sqrt{2}$ is the~$2\times2$ subspace of~$S=1$ spin matrices. The interaction Hamiltonian can be written as
	\begin{equation}\label{Eq_interaction_Hamiltonian}
		H^I = (H-H_0) + e^{iH_0 t} H_{\text{d}} e^{-iH_0 t},
	\end{equation}
	To get a more concise expression, we take
	\begin{equation}
		H_0 = E_1\sigma_{11} + E_2\sigma_{22} + E_6\sum_{i=5}^{8}\sigma_{ii},
	\end{equation}
	thus the second term in~Equation~(\ref{Eq_interaction_Hamiltonian}) becomes
	\begin{equation}\label{Eq:AppA_Hd_int}
		\begin{aligned}
			e^{iH_0 t} H_{\text{d}} e^{-iH_0t} = & \left(e^{-i\omega_p t} \sum_{i=5}^{8} \chi_{1i} \sigma_{1i} + e^{-i\omega_s t} \sum_{i=6,7} \chi_{2i} \sigma_{2i} + h.c.\right) \\
			&\times \left(\Omega_{p}\cos (\omega_{p} t)+\Omega_{s}\cos (\omega_{s} t)\right),
		\end{aligned}
	\end{equation}
	where $\chi_{i j}=\left\langle\psi_{i}|\hat{V}| \psi_{j}\right\rangle$, $\hat{V}=\sqrt{2}\hat{S}_{x}$. Notice here $\chi_{25} = \chi_{52} = \chi_{28} = \chi_{82}=0$ so they are not shown in Equation (\ref{Eq:AppA_Hd_int}). After using the rotating-wave approximation by removing the rapid oscillation terms with frequencies of $2\omega_{p(s)}$ and $\omega_{p(s)}\pm\omega_{s(p)}$ we finally get the final interaction Hamiltonian, as	
	\begin{equation}\label{Eq_App:RWA_Hamiltonian}
		\begin{aligned}
			H^I_{\text{RWA}}=&\frac{\Omega_{p}(t)}{2}\left(\chi_{51} \hat{\sigma}_{51}+\chi_{61} \hat{\sigma}_{61}+\chi_{71} \hat{\sigma}_{71}+\chi_{81} \hat{\sigma}_{81}+\text { h.c. }\right)\\
			&+\frac{\Omega_{s}(t)}{2}\left(\chi_{62} \hat{\sigma}_{62}+\chi_{72} \hat{\sigma}_{72}+\text { h.c. }\right)+(H - H_0).
		\end{aligned}
	\end{equation}

    \section{Hamiltonian with field detuning and amplitude bias}\label{AppSec: H_delta}
    When considering frequency detuning $\delta$ and amplitude bias $\kappa$ of the control fields, the driving Hamiltonian can be represented as
    \begin{equation}
    	H_{\text{d}} = \sqrt{2} (1+\kappa)\left[ \Omega_{p} \cos \left(\omega_{p}' t\right)+\Omega_{s} \cos \left(\omega_{s}' t\right)\right] \hat{S}_{x},
    \end{equation}
    where $\omega_{p}' = E_6-E_1+\delta$, $\omega_{s}' = E_6-E_2+\delta$. Taking $H_0$ as
    \begin{equation}
    	H_0 = E_1\sigma_{11} + E_2\sigma_{22} + (E_6+\delta)\sum_{i=5}^{8}\sigma_{ii},
    \end{equation}
    we have
    	\begin{equation}
    	\begin{aligned}
    		e^{iH_0 t} H_{\text{d}} e^{-iH_0t} = & \left(e^{-i\omega_p' t} \sum_{i=5}^{8} \chi_{1i} \sigma_{1i} + e^{-i\omega_s' t} \sum_{i=6,7} \chi_{2i} \sigma_{2i} + H.c.\right) \\
    		&\times \left[(1+\kappa)\Omega_{p}\cos (\omega_{p}' t)+(1+\kappa)\Omega_{s}\cos (\omega_{s}' t)\right].
    	\end{aligned}
    \end{equation}
    Again we neglect the rapid oscillation terms with frequencies of $2\omega_{p(s)}'$ and $\omega_{p(s)}'\pm\omega_{s(p)}'$, and the interaction Hamiltonian becomes
    \begin{equation}
    	\begin{aligned}
    		H^I_{\text{RWA}}=&\frac{(1+\kappa)\Omega_{p}(t)}{2}\left(\chi_{51} \hat{\sigma}_{51}+\chi_{61} \hat{\sigma}_{61}+\chi_{71} \hat{\sigma}_{71}+\chi_{81} \hat{\sigma}_{81}+\text { h.c. }\right)\\
    		&+\frac{(1+\kappa)\Omega_{s}(t)}{2}\left(\chi_{62} \hat{\sigma}_{62}+\chi_{72} \hat{\sigma}_{72}+\text { h.c. }\right)+\sum_{i=3,4} E_{i} \hat{\sigma}_{ii}\\
    		&+\sum_{i=5}^{8} (E_{i}-E_6-\delta) \hat{\sigma}_{ii}.
    	\end{aligned}
    \end{equation}
	Using this Hamiltonian and according to Equation (\ref{Eq:masterEq}) we can obtain the density matrix $\rho(T,\delta,\kappa)$ at time $T$. Similar to Equation (\ref{Eq:F}), the fidelity with detuning $\delta$ and amplitude bias $\kappa$ can be represented as
	\begin{equation}
		f\left(\delta, \kappa\right) = \text{Tr}(\bra{\psi_{2}}\rho(T,\delta,\kappa)\ket{\psi_{2}}), 
	\end{equation}
	and the average fidelity under different $\delta$ and $\kappa$ can be calculated by Equation (\ref{Eq:F_ave}).

	\section{Pulse Calibrations}\label{AppSec: Experiment}
	
	The method given in the paper for experimentally determining the Rabi frequencies for different pulse shapes is performed by using a linear relationship between the Rabi frequencies and the MW amplitudes from an arbitrary waveform generator~(AWG)~\cite{scheuer2014precise}. This relationship holds due to the small ratio between the transition frequency in the Gigahertz range~(2.9 GHz) and the Rabi frequency in few tens MHz range. Through the pulse calibrations in the experiment we use our in-built signal processing and analysis software~QUDI~\cite{binder2017qudi}. 
	
\begin{figure}[] 
		\centering 
		\includegraphics[width = 5.5cm]{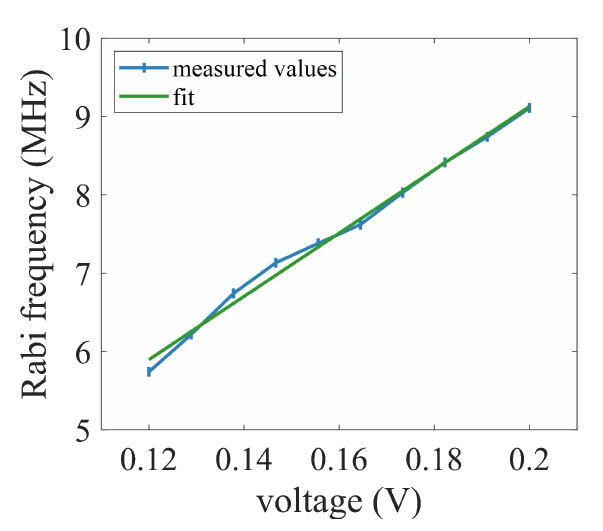}
		\caption{Calibration curve between the input voltage from AWG and the measured Rabi frequency from a single NV experiment. The fit values are $a_{\Omega}=40.4 \pm1.2$~MHz/V, and $b_{\Omega}= 1.0\pm0.2$~V.}
		\label{fig:AWG-Rabi}
	\end{figure}

\begin{figure}[hp]
		\centering
		\includegraphics[width = 5.5cm]{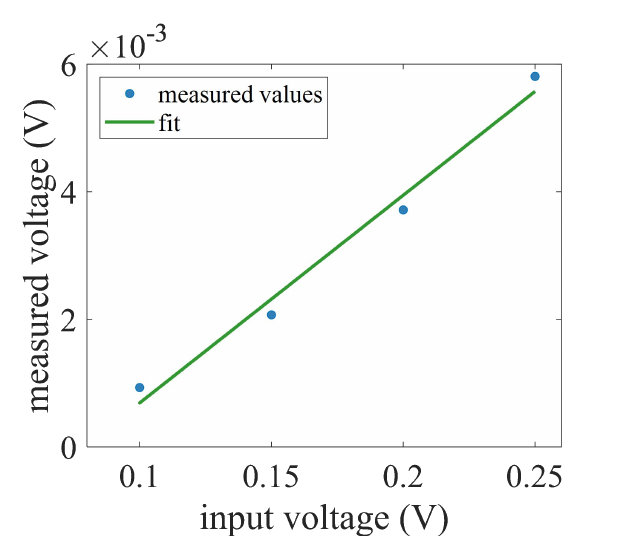}
		\caption{Calibration curve between the input voltage from AWG and the measured voltage from oscilloscope. The fit values are $a_{\text{osc-AWG}}=0.016\pm0.003$ ~V$_{\text{osci}}$/V$_{\text{AWG}}$, and $b_{\text{osc-AWG}} = -0.0026 \pm0.0006$~V$_{\text{osci}}$.}
		\label{fig:osci-AWG}
	\end{figure}

\begin{figure}[hp]
		\centering
		\includegraphics[width = 5.5cm]{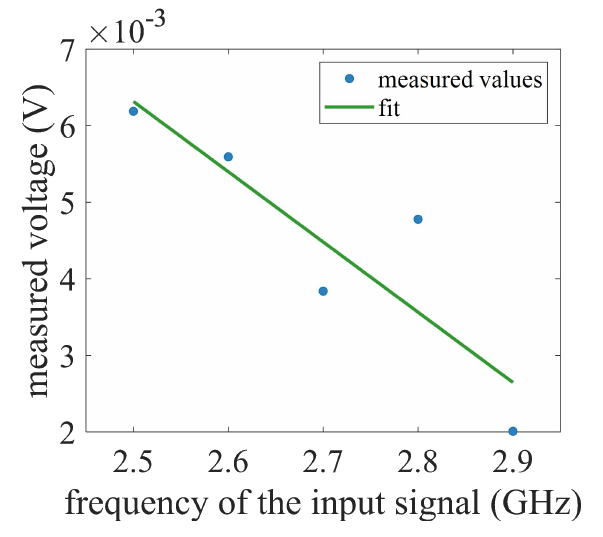}
		\caption{Calibration curve for the relation between the frequency of the input signal and the measured voltage from the oscilloscope~(the input voltage of~$230$~mV). The parameters are $a_{V-f}= -0.0092\pm0.0028$~V/GHz, and~$b_{V-f}=0.029\pm 0.008$~V.}
		\label{fig:freq_calib_fix_ampli}
	\end{figure}

	For further translation of the pulsed signal from the oscilloscope voltage signal to the Rabi frequency~(see Figure~\ref{fig:AWG-Rabi}), the same linear matching procedure is performed between the voltage from the AWG and the amplified signal amplitude measured by the oscilloscope~(see Figure~\ref{fig:osci-AWG}). Also, we perform an analysis on the relationship between the signal at constant AWG voltage and different frequencies of the MW signals. This is due to fact that the oscilloscope has a certain bandwidth, where the amplitude varies at different frequencies (here, the oscilloscope used has an operating bandwidth up to~300~MHz). After the critical value it has a certain damping, which has a linear behaviour at the higher frequency range~(see Figure~\ref{fig:freq_calib_fix_ampli}). 
	
	Here, the relation between the measured voltage of the oscilloscope~$V_{\text{osci}}$ and the voltage inserted in the AWG~$V_{\text{AWG}}$ is obtained by the following linear equation, 
	\begin{equation*}
		V_{\text{osci}} = a V_{\text{AWG}} + b,
	\end{equation*}
	with the scaling factor~ $a$ and an offset~$b$, obtained by the linear regression. We use the same procedure to establish the relations among the Rabi frequency~$\Omega$, the voltage~$V_{AWG}$, and the signal frequency~$f$. From these previous measurements the comparison between the experimental data and the simulated data are done (see Figure 4).
	
	\newpage
	
	\begin{figure*}[hp]\label{Total_pulses2}
		\centering
		\includegraphics[width = 18cm]{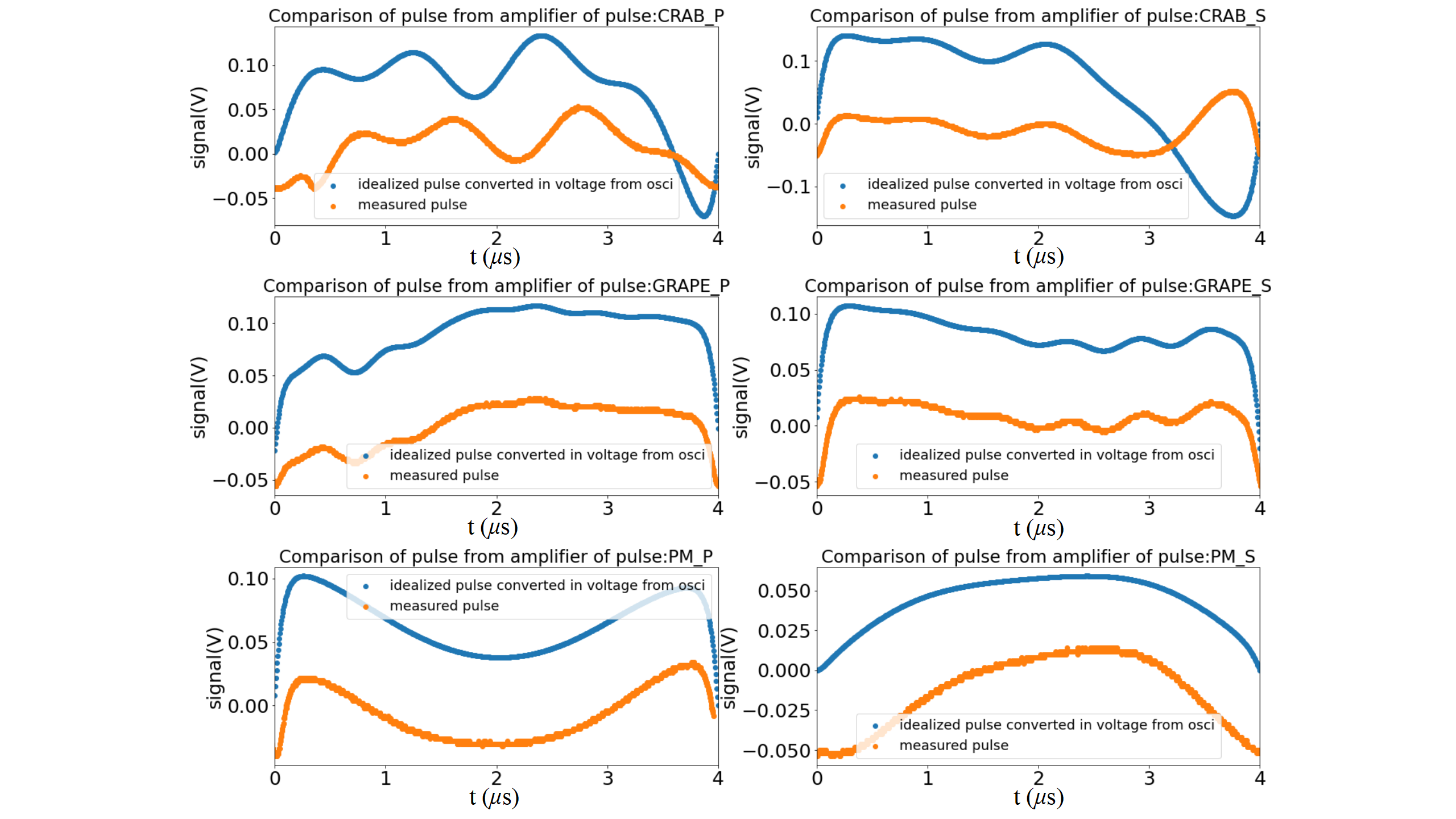}
		\caption{Comparison of the pulse shapes. For clarity, the scale is pre-determinedly chosen to simply compare the pulse shapes from the numerically-obtained pulses and the measured ones. For the case of~$\Omega_\text{s}$ obtained via the CRAB method at time of~$3$ $\mu$s, the amplitude goes in the upper direction according to the pulse envelope of the electronical signal.}
		\label{fig:Comp Pulses}
	\end{figure*}

	\bibliography{CFSbib}
	
\end{document}